\documentclass[aps,prl,twocolumn,showpacs,superscriptaddress]{revtex4-1}

\usepackage{graphics}
\usepackage{graphicx}
\usepackage{dcolumn}
\usepackage{bm}
\usepackage{color}
\usepackage{amssymb,amsmath}

\begin{document}

\title{Shock-induced  $\mathcal{PT}$-symmetric potentials in gas-filled photonic crystal fibers}
\author{Mohammed F. Saleh}
\affiliation{Max Planck Institute for the Science of Light, G\"{u}nther-Scharowsky str. 1, 91058 Erlangen, Germany}
\author{Andrea Marini}
\affiliation{Max Planck Institute for the Science of Light, G\"{u}nther-Scharowsky str. 1, 91058 Erlangen, Germany}
\author{Fabio Biancalana}
\affiliation{Max Planck Institute for the Science of Light, G\"{u}nther-Scharowsky str. 1, 91058 Erlangen, Germany}
\affiliation{School of Engineering and Physical Sciences, Heriot-Watt University, EH14 4AS Edinburgh, UK}
\date{\today}

\begin{abstract}
We have investigated the interaction between a strong soliton and a weak probe with certain configurations that allow optical trapping in gas-filled hollow-core photonic crystal fibers in the presence of the shock effect. We have shown theoretically and numerically that the shock term can lead to an unbroken parity-time $ \left( \mathcal{PT}\right)  $ symmetry potential in these kinds of fibers. Reciprocity breaking, a remarkable feature of the $ \mathcal{PT}  $ symmetry, is also demonstrated numerically. Our results will open different configurations and avenues for observing $ \mathcal{PT} $-symmetry breaking in optical fibers, without the need to resort to cumbersome dissipative structures.
\end{abstract}
\maketitle

One of the postulates of ordinary quantum mechanics is that Hermitian operators are associated to physical observables, which are accompanied by a spectrum of real eigenvalues \cite{Landau77}. Surprisingly, Bender \textit{et al.} in the late nineties have found a class of \textit{non-Hermitian} Hamiltonian operators that can also exhibit entirely real spectra, provided that these operators satisfy the parity-time $ \left( \mathcal{PT}\right)  $ symmetry \cite{Bender98,Bender99}. Specifically, a Hamiltonian operator is said to be $  \mathcal{PT}  $ symmetric, if $ U\left( x\right)=U^{*}\left(- x\right)  $, where $ U$ is the potential, and $ x $ is the position. Therefore, the real part of the potential must be even, while its imaginary part should be odd. Interestingly, this class of operators are characterized by the existence of a certain threshold above which the $  \mathcal{PT}  $ symmetry is spontaneously broken, and the entire spectrum of eigenvalues becomes complex. The $  \mathcal{PT}  $ symmetry has been investigated and explored theoretically \cite{Ganainy07,Musslimani08a,Musslimani08b,Longhi09, Sukhorukov10,Chong11,Szameit11,Longhi11,Alexeeva12,Sheng13} and demonstrated experimentally \cite{Klaiman08,Guo09,Rueter10,Regensburger12} in a number of different optical settings that exhibit gain and loss processes simultaneously. The structure complexity is the most common feature in these microstructures.

After the invention of the hollow-core (HC) photonic crystal fibers (PCFs) \cite{Russell03,Russell06}, there is currently a great motivation to demonstrate the $  \mathcal{PT}  $ symmetry in the most successful waveguide of all, the optical fiber. Usually, the Raman nonlinear contribution is dominant in solid-core fibers, and can deteriorate this interesting phenomenon. However, HC-PCFs based on a Kagome lattice have extended the field of linear and nonlinear fiber optics well beyond the interaction of light with solid media \cite{Travers11}.  A HC-PCF can be filled with Raman-inactive gases such as noble gases, allowing unprecedented possibilities to study `pure' nonlinear effects without the `disturbance' of the Raman effect. In recent years, the study of HC-PCFs filled with noble gases has led to the demonstration and the prediction of interesting unexpected phenomena such as high harmonic generation \cite{Heckl09}, deep UV generation \cite{Joly11}, soliton self-frequency blue shift \cite{Chang11,Hoelzer11b,Saleh11a,Saleh11b,Chang13}, asymmetric self phase modulation, and universal modulational instability \cite{Saleh12}.

In this Rapid Communication, we propose the HC-PCF as an alternative structure in order to observe the $  \mathcal{PT}  $ symmetry in optics rather than using cumbersome microstructures in which gain and loss must be carefully balanced. Very recently, we have introduced a system to observe optical trapping between two ultrashort pulses in HC-PCFs filled by a noble gas due to solely the cross phase modulation effect (XPM) \cite{Saleh13}. The system is Raman and plasma free. The two pulses have different scale of intensities, the strong `pump' pulse introduces a potential well that traps the weak `probe' pulse inside it. The two pulses must have the same frequency to fulfil the group-velocity matching condition, and different circular polarization states for a clear monitoring at the fiber output. In order to eliminate birefringence-induced coupling between the pulses, the fiber core should be perfectly symmetric. In this case, the propagation of the two pulses in a lossless medium are governed by the following set of normalized coupled nonlinear Schr\"{o}dinger equations \cite{Agrawal07},
\begin{equation}
\begin{array}{l}
i\,\partial_{\xi}\psi_{1}+\frac{1}{2}\partial_{\tau}^{2}\psi_{1}+\left|\psi_{1}\right|^{2}\psi_{1}+i\, \tau_{\rm sh}\,\partial_{\tau}\left( \left|\psi_{1}\right|^{2}\psi_{1}\right)=0 ,\\
i\,\partial_{\xi}\psi_{2}+\frac{1}{2}\:\partial_{\tau}^{2}\psi_{2}+2\left|\psi_{1}\right|^{2}\psi_{2}+i\, 2\, \tau_{\rm sh}\,\partial_{\tau}\left( \left|\psi_{1}\right|^{2}\psi_{2}\right)=0 ,
\end{array}
\end{equation}
where $ \psi_{1} $ and $ \psi_{2} $ are the complex envelopes of the pump and the probe, respectively, $\xi$ is the longitudinal coordinate along the fiber, $ \tau $ is the time coordinate, $ \partial $ is the partial derivative, and $ \tau_{\rm sh} $ is the normalized shock coefficient. The effect of higher-order dispersion coefficients, and the nonlinearity of the probe are neglected.

\begin{figure}
\includegraphics[width=8.6cm]{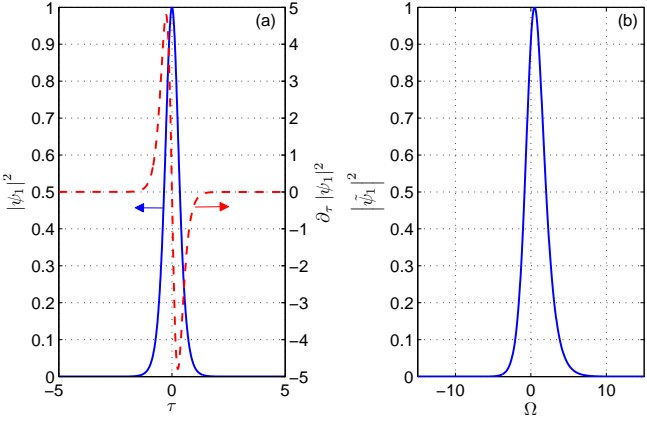}
\caption{(Color online). Solid blue curves represent (a) the temporal and (b) the spectral profile of the pump intensity, respectively. $ \Omega $ is the normalized frequency shift, $ N =2.5 $, and $ \tau_{\rm sh}=0.1137 $. These parameters are used in all the simulations shown in this rapid communication, except in Fig. \ref{Fig1} where $ N $ is an independent variable. The time derivative of the temporal pulse intensity is shown by the dashed red curve in (a). 
\label{Fig0}}
\end{figure}

The governing equation of the propagation of the pump is found to be integrable, and has a family of soliton solutions \cite{Anderson83,Ping00}. The fundamental solution has a form similar to the Schr\"{o}dinger soliton, but it is chirped, 
\begin{equation}
\psi_{1}\left( \xi,\tau\right) =A\, \mathrm{sech}\:\theta \frac{B^{*}+1+\left( B^{*}-1\right)\mathrm{tanh}\:\theta}{\left[ B+1+\left( B-1\right)\mathrm{tanh}\:\theta\right] ^{2}}\,e^{iN^{2}\xi/2},
\label{soliton}
\end{equation}
where $ \theta=N \tau $, $ A=2N\left(1+N^{2}\tau_{\rm sh}^{2} \right) ^{-1/4} $, $ B=\left(1+iN\tau_{\rm sh} \right)A^{2}/\left( 4N^{2}\right)  $, and $ N $ is the soliton order. The temporal and the spectral profiles of the pump intensity are depicted in panels (a) and (b) of Fig. \ref{Fig0}, respectively. It is clearly shown that the temporal dependence is symmetric, while the spectral dependence is asymmetric.

The probe propagates linearly, since its nonlinearity is negligible. Its governing equation can be written as 
\begin{equation}
\begin{array}{rc}
 i\,\partial_{\xi}\psi_{2}+i\,2\, \tau_{\rm sh} \left|\psi_{1}\right|^{2}\,\partial_{\tau}\psi_{2}+\frac{1}{2}\:\partial_{\tau}^{2}\psi_{2}&  \vspace{1mm}\\ 
 +2\left[ \left|\psi_{1}\right|^{2}+i\, \tau_{\rm sh}\,\partial_{\tau} \left|\psi_{1}\right|^{2}\right] \psi_{2}& =0.
\end{array} \label{eq2}
\end{equation}
Hence, the shock term introduces two terms proportional to $\tau_{\rm sh}$, as shown in Eq. (\ref{eq2}). The first term results in additional time-dependent group velocity, while the second term modifies the trapping potential by an imaginary component. Introducing the following phase transformation
\begin{equation}
\psi_{2}=\tilde{\psi}\:\exp\left[-i\,2\,\tau_{\rm sh}\displaystyle\int\left|\psi_{1}\right|^{2} dt \right],
\end{equation}
Eq. (\ref{eq2}) can be simplified to
\begin{equation}
i\,\partial_{\xi}\tilde{\psi}+\frac{1}{2}\:\partial_{\tau}^{2}\tilde{\psi}
 +\left[ 2\left|\psi_{1}\right|^{2}+2\, \tau_{\rm sh}^{2}\, \left|\psi_{1}\right|^{4}+i\, \tau_{\rm sh}\,\partial_{\tau} \left|\psi_{1}\right|^{2}\right] \tilde{\psi} =0. \label{eq4}
\end{equation}
Seeking for stationary solutions for the probe $ \tilde{\psi}\left( \xi,\tau\right)=f\left( \tau\right)\exp\left( -iq\xi\right) $, with temporal profile $ f $, and propagation constant $ q $, Eq. (\ref{eq4}) becomes a linear Schr\"{o}dinger equation in time,
\begin{equation}
-\frac{1}{2}\:\partial_{\tau}^{2}f+U\left(\tau \right) \,f=q\,f,
\end{equation}
where $ U= - 2\left|\psi_{1}\right|^{2}-2\, \tau_{\rm sh}^{2}\, \left|\psi_{1}\right|^{4}-i\, \tau_{\rm sh}\,\partial_{\tau} \left|\psi_{1}\right|^{2}$ is the potential well, and $ q $'s are the eigenvalues. \textit{This potential possesses a $\mathcal{PT}  $-symmetry, since its real part is even, while its imaginary part is odd}, see Fig. \ref{Fig0} (a). Note that such potential is only due to the presence of the pump soliton $\psi_{1}$, and  the shock term that is always present in optical fiber systems, and cannot be neglected. In addition, the phase structure of the soliton does not influence the form of the potential, since $U$ depends only on $|\psi_{1}|$. 

\begin{figure}
\includegraphics[width=8.6cm]{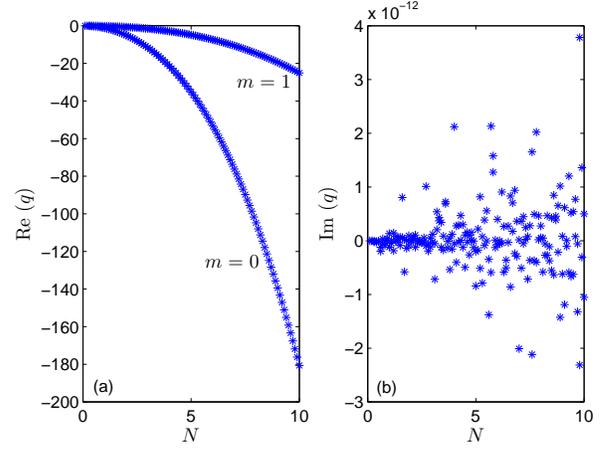}
\caption{(Color online). The soliton-order dependence of (a) the real and (b) the imaginary parts of the propagation constants of the probe in the fundamental, $ m=0 $, and the first-order, $ m=1  $, modes.
\label{Fig1}}
\end{figure}

Similarly to the case in the absence of the shock coefficient \cite{Saleh13}, this potential well has always two localized modes. These modes can be computed numerically using the sparse matrix technique \cite{Benham01}. The dependence of the real and the imaginary parts of the propagation constants on the soliton order are depicted in Fig. \ref{Fig1} for both modes. As shown, the values of the imaginary parts are negligible in comparison to the real parts, as well as they follow a random distribution, which indicates that these values are just numerical errors. In other words, the propagation constants or the eigenvalues are always pure real values, and there is no threshold for breaking the $\mathcal{PT}  $-symmetry. In general, the real part of the potential tries to maintain the symmetry of the probe solution, while the imaginary part would be responsible for the break up of this symmetry. In our case, however, the real part of the potential stays dominant in comparison to the imaginary part, because the shock term affects both parts simultaneously with different powers  of the soliton order $ N $.  This result contradicts a general statement that a linear eigenvalue problem has a zero threshold point or the $\mathcal{PT}  $-symmetry in the corresponding system is always broken \cite{Musslimani08b}.

\begin{figure}
\includegraphics[width=8.6cm]{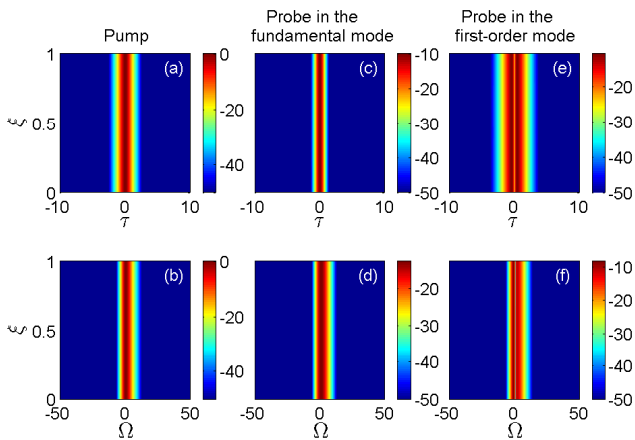}
\caption{(Color online). (a,b) Temporal and spectral evolution of the pump. (c,d) Temporal and spectral evolution of the copropagating probe, when it is in the fundamental mode. (e,f) Temporal and spectral evolution of the copropagating probe, when it is in the first-order mode.
\label{Fig2}}
\end{figure}

The co-propagation of the pump and the probe in either the fundamental or the first-order mode are shown in Fig. \ref{Fig2}. The probe does not suffer from dispersion-induced broadening during propagation due to XPM that traps the probe inside the soliton-induced potential.

\begin{figure}
\includegraphics[width=8.6cm]{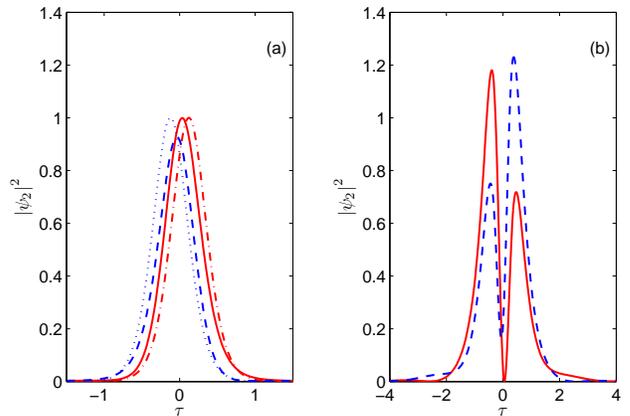}
\caption{(Color online). Temporal profiles of the probe at the fiber output, when it is in (a) the fundamental or (b) the first-order mode. Solid red and dashed blue curves represent the output profiles when the inputs are delayed by $ \Delta\tau=10 $ and $ \Delta\tau=-10 $, respectively. Red dashed-dotted and  blue dotted curves in (a) represent the input probe, when it is delayed by $ \Delta\tau=10 $ and $ \Delta\tau=-10 $, respectively.
\label{Fig3}}
\end{figure}

$ \mathcal{PT} $-symmetric waveguides are characterized by breaking reciprocity even if the symmetry is not broken \cite{Rueter10}. In the space domain, having coupled $ \mathcal{PT} $-symmetric waveguides, the light evolution is not the same when it is launched in one waveguide or the other. Mimicking this experiment in the time domain by introducing a positive or negative delay $ \Delta\tau $ (with the same magnitude) between the pump and the probe, non-mirror symmetry outputs are obtained. Panel (a) of Fig. \ref{Fig3} shows the output temporal profiles of the probe when it is launched in the fundamental mode with $ \Delta\tau=\pm 10 $. As depicted, the two outputs are not a mirror symmetry of each other, unlike the inputs. The reason is due to the shock term that modifies the soliton-induced potential and transforms it from a regular symmetric one to a $ \mathcal{PT} $-symmetric potential. As the shock term increases, this reciprocity breaking increases, even though $ \mathcal{PT} $-symmetry itself is never broken. Panel (b) shows the reciprocity breaking when the probe is launched in the first-order mode.

In conclusion, we have studied the interaction between a strong soliton and a weak probe in a symmetric HC-PCF filled by a noble gas in the presence of the self steepening (shock) effect. The two pulses have the same central frequency, and opposite circular polarization states. The medium is Raman and plasma free. The probe is trapped during propagation because of the soliton-induced potential that has always two localized modes. We have proven theoretically that the shock term modifies the soliton-induced potential, which becomes $ \mathcal{PT} $-symmetric, by introducing an odd imaginary component to the potential. We have found that the eigenvalues of the system are always real, or in other words the $ \mathcal{PT} $-symmetry is always unbroken. Finally, we have demonstrated numerically the reciprocity breaking by introducing a positive or negative time delay (with the same absolute value) between the pump and the probe. We believe that our results will open different configurations and avenues for observing $ \mathcal{PT} $-symmetry breaking in optical fibers, without the need to resort to cumbersome dissipative structures.

This research is funded by the German Max Planck Society for the Advancement of Science (MPG).

\bibliographystyle{apsrev4-1}	

%

\end{document}